\documentclass[sigconf, natbib=false, nonacm]{acmart}

\usepackage{amsmath,amsfonts}
\usepackage{bm}
\usepackage{nicefrac}
\usepackage{siunitx}
\usepackage{graphicx}
\usepackage{xcolor}
\usepackage{subcaption}
\usepackage{siunitx}
    
\usepackage[version=4]{mhchem}    
\usepackage{subcaption}
    
\usepackage{tikz}
\usetikzlibrary{math}
\usetikzlibrary{shapes.geometric,calc}
\usetikzlibrary{arrows.meta, quotes, angles}
\usetikzlibrary{decorations}
\usetikzlibrary{chains,shapes.misc,positioning,scopes}
\usepackage{circuitikz}
\usetikzlibrary{circuits.ee.IEC}
\usetikzlibrary{decorations}
\usetikzlibrary{shadows}
\usepackage{subcaption}
\usepackage{acronym}

\acrodef{CAM}{Chorioallantoic Membrane}
\acrodef{ICG}{Indocyanine Green}
\acrodef{IoBNT}{Internet of BioNanoThings}
\acrodef{MC}{Molecular Communications}
\acrodef{DED}{Day of Embryonic Development}
\acrodef{TX}{transmitter}
\acrodef{RX}{receiver}
\acrodef{nRX}{natural receiver}
\acrodef{sRX}{synthetic receiver}
\acrodef{ISI}{Inter Symbol Interference}
\acrodef{ROI}{Region of Interest}

\newcommand{\veff}{v_\mathrm{eff}}

\newcommand{\Deff}{D_\mathrm{eff}}
\newcommand{\Leff}{L_\mathrm{eff}}

\AtBeginDocument{%
  }

\setcopyright{acmlicensed}
\copyrightyear{2024}
\acmYear{2024}
\acmDOI{XXXXXXX.XXXXXXX}

\acmConference[ACM NanoCom 2024]{11th ACM International Conference on Nanoscale Computing and Communication}{October 28--30, 2024}{Milan, Italy}
\acmISBN{978-1-4503-XXXX-X/18/06}

\RequirePackage[
  datamodel=acmdatamodel,
  style=acmnumeric,
  ]{biblatex}

\newcommand\varcal[1]{\text{\usefont{OMS}{cmsy}{m}{n}#1}}

\begin{document}
\setlength{\baselineskip}{10.6pt}
\title{The Chorioallantoic Membrane Model: A 3D \textit{in vivo} Testbed \\ for Design and Analysis of MC Systems}

\author{Maximilian Schäfer\textsuperscript{1}, Andreas Ettner-Sitter\textsuperscript{2}, Lukas Brand\textsuperscript{1}, Sebastian Lotter\textsuperscript{1}, \\Fardad Vakilipoor\textsuperscript{1}, Thiha Aung\textsuperscript{3}, Silke Haerteis\textsuperscript{2}, and Robert Schober\textsuperscript{1}}
\affiliation{%
  \institution{\textsuperscript{1}Friedrich-Alexander-Universit\"{a}t Erlangen-N\"urnberg, Erlangen, Germany}
  \country{}
}
\affiliation{%
  \institution{\textsuperscript{2} University of Regensburg, Regensburg, Germany}
\country{}
}
\affiliation{%
  \institution{\textsuperscript{3}Deggendorf Institute of Technology, Deggendorf, Germany}
\country{}
}

\renewcommand{\shortauthors}{Schäfer et al.}

\begin{abstract}
\acf{MC} research is increasingly focused on applications within the human body, such as health monitoring and drug delivery. These applications require testing in realistic and living environments. Thus, advancing experimental \ac{MC} research to the next level requires the development of \textit{in vivo} experimental testbeds. In this paper, we introduce the \acf{CAM} model as a versatile 3D \textit{in vivo} \ac{MC} testbed. The \ac{CAM} is a highly vascularized membrane formed in fertilized chicken eggs and has gained significance in various research fields, including bioengineering, cancer research, and drug development. Its versatility, reproducibility, and realistic biological properties make the \ac{CAM} model perfectly suited for next-generation \ac{MC} testbeds, facilitating the transition from  proof-of-concept systems to practical applications. We provide a comprehensive introduction to the \ac{CAM} model, its properties, and its applications in practical research. Additionally, we present a characterization of the \ac{CAM} model as an MC system. As a preliminary experimental study, we investigate the distribution of fluorescent molecules in the closed-loop vascular system of the \ac{CAM} model. We also derive an approximate analytical model for the propagation of molecules in closed-loop systems, and show that the proposed model is able to approximate molecule propagation in the \ac{CAM} model.
\end{abstract}





\maketitle

\section{Introduction}

Synthetic \ac{MC} is an emerging interdisciplinary research field at the intersection of life sciences and engineering \cite{Nakano2011}. \ac{MC} naturally occurs in biological systems and is envisioned to enable communication between synthetic nanomachines and biological entities. In recent years, numerous potential applications of MC have been identified across various sectors, including industry, environmental monitoring, and medicine \cite{Farsad2016,Felicetti2016}. 

Despite the recent increase of experimental work, most \ac{MC} research remains theoretical. Bridging the gap between theoretical concepts and their practical application is a significant challenge \cite{Lotter2023b}. 
In the medical sector, \ac{MC} often focuses on in-body applications, necessitating the validation of the concepts and technologies in realistic \textit{in vivo} environments. However, many promising MC technologies are highly invasive and disruptive, making animal or human testing very difficult and currently almost impossible. 

\begin{figure}[t]
    \vspace*{-0.1cm}
    \centering
    \includegraphics[width=0.6\linewidth]{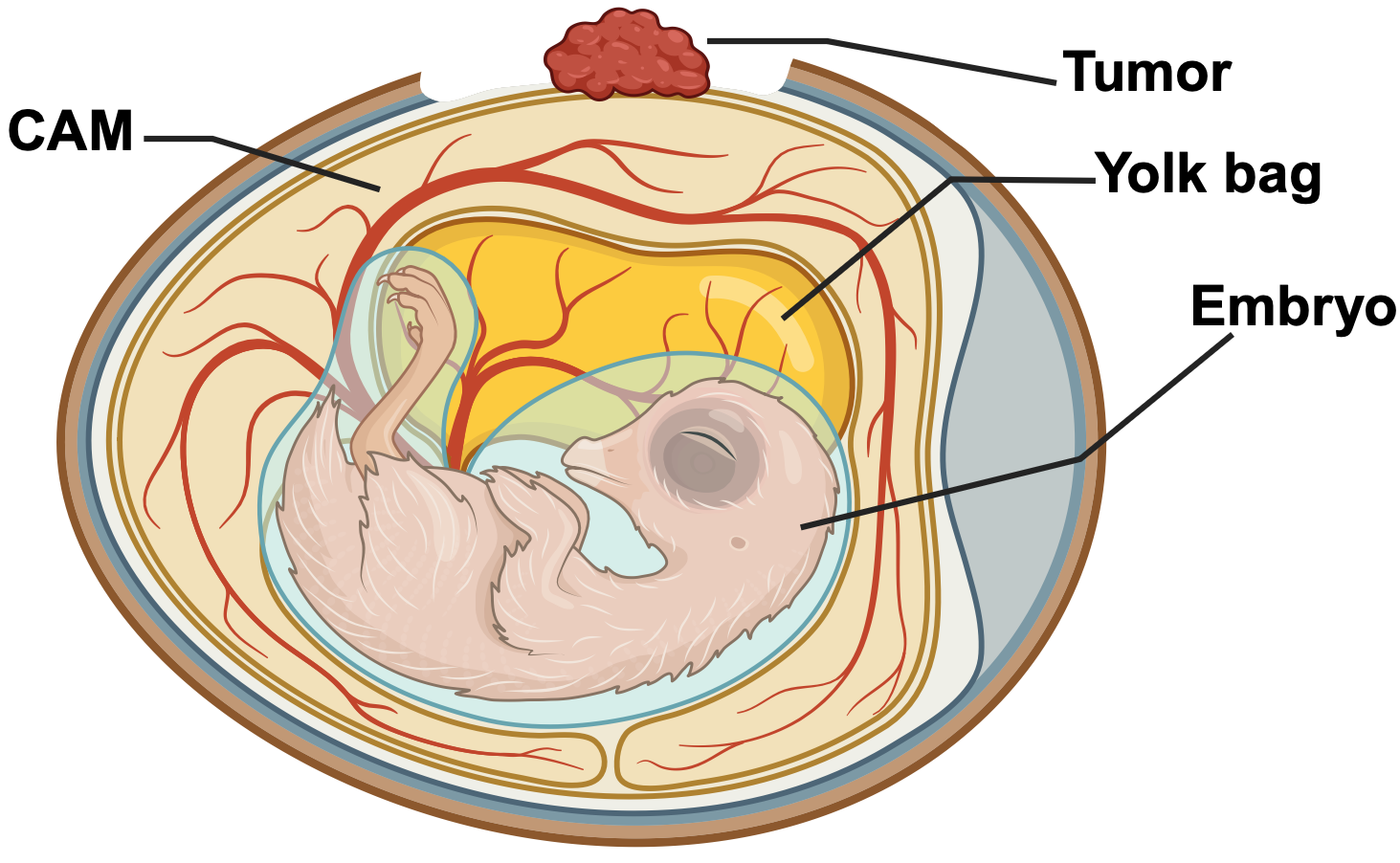}
    \vspace*{-0.45cm}
    \caption{\small Schematic illustration of the CAM model consisting of the chicken embryo, the yolk bag, and the \ac{CAM} established. Furthermore, a tumor is engrafted on the CAM (Created with \url{BioRender.com}).}
    \label{fig:cam-schematic}
    \vspace*{-0.68cm}
\end{figure}
Experimental work on in-body \ac{MC} is often focused on concepts for exchanging information between in-body devices, such as mobile nanosensors patrolling the cardiovascular system, and devices for the controlled delivery of drug molecules, and their experimental validation. Hereby, most testbeds established so far employ a tube-like channel (e.g., a pipe or a microfluidic chip) as a rough approximation of the cardiovascular system, where the signaling molecules are injected via a syringe or a valve \cite{Farsad2017,Pan2022,Unterweger2018,Tuccitto2017,Wang2020}. 
While the channel structures and transmission mechanisms are mostly similar, these testbeds differ in the choice of the signaling molecule and thus also in the choice of the receiver structures. For example, in \cite{Unterweger2018}, superparamagnetic iron oxide nanoparticles are injected into a tube and detected by a susceptometer. In \cite{Tuccitto2017}, fluorescent nanoparticles are employed as signaling molecules which are detected by a photodiode, and \cite{Pan2022} uses color pigments as signaling molecules detected by a color sensor. 
Despite this significant progress in experimental \ac{MC} research in the last few years, the influence of realistic and living - \textit{in vivo} - environments on the proposed \ac{MC} system designs has not been investigated, yet. To enable the practical deployment of synthetic MC systems in the future, it is indispensable to elevate experimental \ac{MC} research to the next level by developing \textit{in vivo} testbeds. 

In this paper, we present a 3D \textit{in vivo} \ac{MC} testbed based on the \acf{CAM} model, which provides an accessible model of a cardiovascular system including blood circulation and organs (see Fig.~\ref{fig:cam-schematic}). 
The \ac{CAM} is formed in fertilized chicken eggs as a highly vascularized extraembryonic membrane that functions as a respiratory organ \cite{Mapanao21, Valdes2002}. Human cells or tissues can be engrafted onto the \ac{CAM}, e.g., to study the effect of potential therapeutics. The \ac{CAM} model is well established in different research fields and was originally used for angiogenesis and anti-angiogenic therapeutic approaches. In recent years, however, it has gained more and more importance in many different fields, such as bioengineering, transplant biology, cancer research, and drug development \cite{Feder2020}. Moreover, the \ac{CAM} model fulfills the 3R principles for the reduction, refinement, and replacement of animal models for research purposes and does not require an ethic vote in most countries.

As the \ac{CAM} model represents a simple and accessible \textit{in vivo} model of a cardiovascular system including blood circulation and organs in the chick embryo, it is perfectly suited as a next generation \ac{MC} testbed. Therefore, in this paper, we propose the \ac{CAM} model as a versatile and realistic model of an \textit{in vivo} system for the validation and optimization of \ac{MC} technologies, to enable the transition from simple proof-of-concept MC systems towards practical applications \cite{Ettner2024}.
The contributions of this paper can be summarized as follows:
\begin{itemize}
    \setlength\itemsep{0em}
    \item For the first time, we propose the \ac{CAM} model as a versatile and accessible \textit{in vivo} \ac{MC} testbed, and discuss its capabilities for the analysis and design of \ac{MC} systems. 
    \item We propose an analytical model for the diffusion and advection of particles in dispersive closed-loop systems, such as the \ac{CAM} model. 
    \item Based on fluorescence measurements of the distribution of molecules in the \ac{CAM} model, we show that the proposed analytical model is able to approximate the propagation of molecules in the \ac{CAM} model.  
\end{itemize}

The remainder of the paper is organized as follows: In Section~\ref{sec:CAM}, we introduce the \ac{CAM} model, its topology and applications. Moreover, we characterize the \ac{CAM} model from an \ac{MC} point of view. In Section \ref{sec:particleDis}, we propose an analytical model for the dispersion of molecules in closed-loop systems. In Section \ref{sec:exp}, we apply the proposed model to approximate the distribution of molecules in the \ac{CAM} model and compare it to measurements. Finally, we draw our main conclusions and propose topics for future work in Section~\ref{sec:conc}. 

\begin{figure}[t]
    \centering
    \includegraphics[width=0.8\linewidth]{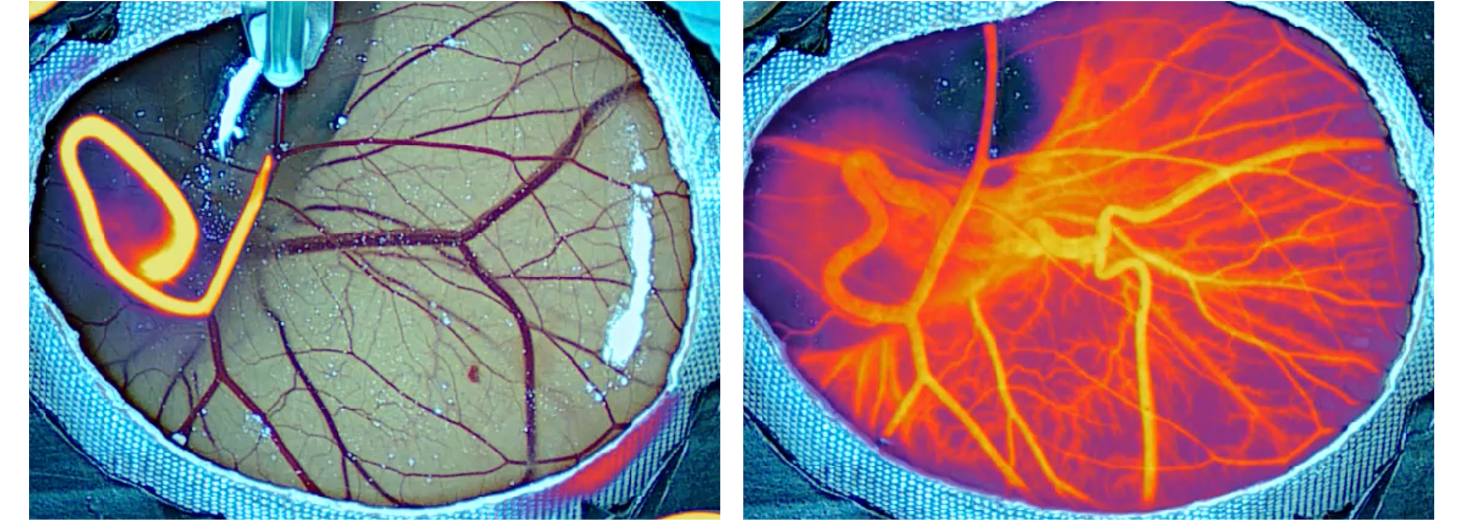}
    \vspace*{-0.4cm}
    \caption{\small Snapshots of the distribution of fluorescent molecules in the vascular system of the \ac{CAM} model directly after injection via a syringe (left) and after a few seconds fully distributed (right). 
    }
    \label{fig:icg-injection}
    \vspace*{-0.6cm}
\end{figure}

\vspace*{-0.1cm}
\section{Chorioallantoic Membrane Model}
\label{sec:CAM}

The \ac{CAM} model is a 3D \textit{in vivo} model bridging the gap between preclinical research and clinical trials \cite{Mapanao21, Valdes2002}. As the \ac{CAM} model has not been considered before in the \ac{MC} community, in the following, we provide a short overview of the components of the \ac{CAM} model, describe the main parameters of the developing vascular system, and explain the growing importance of the \ac{CAM} model in tumor and drug distribution studies. Finally, we characterize the \ac{CAM} model as an \ac{MC} system. 

\vspace*{-0.2cm}
\subsection{Topology}
\label{sec:topol}
As shown in Fig.~\ref{fig:cam-schematic}, the \ac{CAM} model includes three main elements: the chick embryo, the yolk bag, and the \ac{CAM} surrounding the other elements. 

The \ac{CAM} itself is an extra-embryonic membrane that is formed by the mesodermal fusion of allantois and chorion within the third and tenth \ac{DED}. During this period, it functions primarily as the respiratory organ of the embryo \cite{kundekova2021chorioallantoic, nowak2014chicken}. The \ac{CAM} is comprised of a vast array of blood vessels and a dense capillary network. Due to the high level of vascularization, it has been widely utilized as a platform for \textit{in vivo} studies of vascular development and angiogenesis \cite{Guerra2021}. The vascular structure developing in the \ac{CAM} is completely connected to and supplied by the embryo’s vascular system and -- depending on the day of evolution -- also to the organs of the embryo. The yolk bag is needed to nourish the embryo and takes over the function of organs such as the liver, until these organs have fully developed in the embryo.
Therefore, the \ac{CAM} model represents a realistic \textit{in vivo} model of a circulatory system providing a highly complex branched vascular system as well as several natural environmental effects such as various clearance mechanisms caused by organs.  


\begin{figure}[t]
    \centering
    \includegraphics[width=0.8\linewidth]{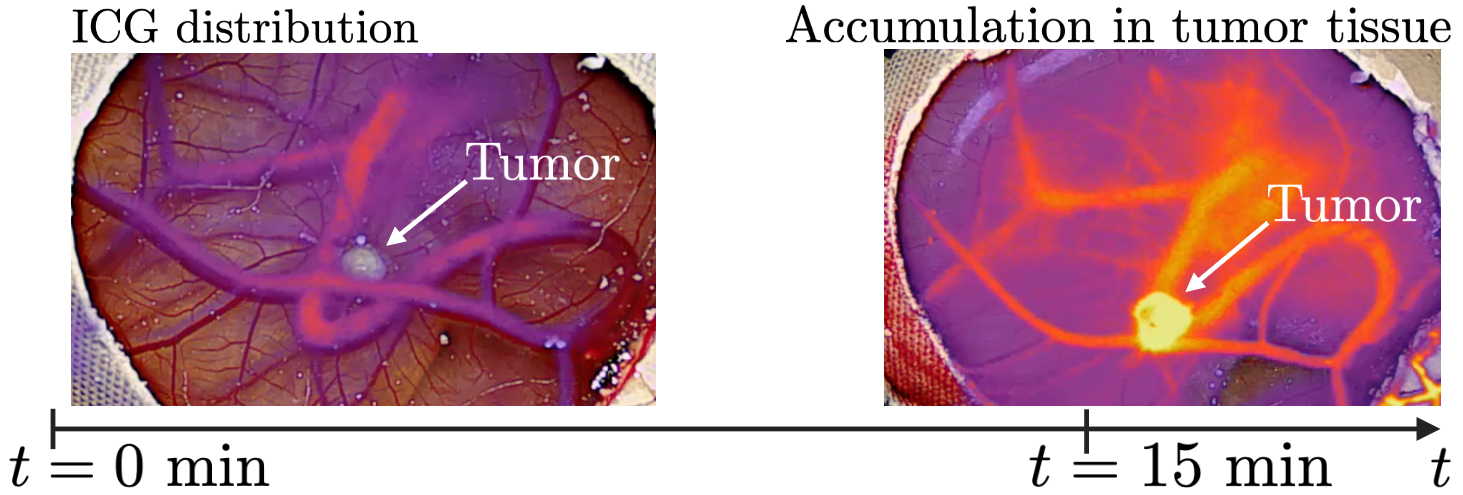}
    \vspace*{-0.4cm}
    \caption{\small \ac{CAM} model for tumor and distribution studies. Left: \ac{ICG} injection into \ac{CAM} vessels at $t = 0~\si{\minute}$ and distribution in the vascular system. Right: \ac{ICG} is accumulated in tumor tissue after $\sim \!15~\si{\minute}$ \cite{Ettner2024}.}
    \label{fig:cam-drug-study}
    \vspace*{-0.5cm}
\end{figure}

\vspace*{-0.2cm}
\subsection{Particle Propagation}
Characterizing the influence of diffusion, flow, and vascularization on the propagation of particles inside the \ac{CAM} model is not straightforward. One of the biggest challenges is in fact that the \ac{CAM} is a constantly developing system \cite{Makanya2016, Guerra2021}. Depending on the \ac{DED}, the vascular system in the \ac{CAM} and in the embryo is changing significantly and therefore, the blood volume, average volume flux, and also other parameters depend on the \ac{DED} \cite{Tazawa1977}. Also, the effective local blood flow velocity in the \ac{CAM} model strongly depends on the \ac{DED} and possibly exhibits a large variance among different eggs. In \cite{Kloosterman2014}, it has been shown that the flow profile is laminar and measured time-averaged mean velocities range from $1~\si{\micro\meter\per\second} - 1~\si{\milli\meter\per\second}$ for vessel diameters ranging from $25 - 500~\si{\micro\meter}$. In \cite{Maibier2016}, it has been additionally shown that the topology of the \ac{CAM} can be divided into feeding arterioles and draining venules having larger diameters, and secondary arterioles and venules with diameters normally distributed with means of $46.3~\si{\micro\meter}$ and $49~\si{\micro\meter}$, respectively. Moreover, the authors of \cite{Maibier2016} showed that the measured effective blood flow velocities are increasing lineary with the vessel diameter \cite[Fig.~4]{Maibier2016}. The influence of diffusion inside the \ac{CAM} model depends on the properties of the particles injected and the occurring chemical reactions, e.g., the affinity and binding of molecules to serum proteins such as albumin \cite{Weixler23}. 

Depending on the \ac{DED}, the yolk bag and the organs of the chick embryo can have strong influence on the propagation of particles in the \ac{CAM} model, as they introduce additional clearance and accumulation effects. In the first \acp{DED}, the yolk bag takes over the functions of the liver to filter toxic particles, and therefore particles entering the yolk bag will accumulate with a certain probability. In later \acp{DED}, organs in the embryo (e.g., brain and liver) start to develop and particles accumulate there, too \cite{Ettner2024}. 
Moreover, tumor tissue can be engrafted on the \ac{CAM} (see Fig.~\ref{fig:cam-drug-study}), where particles can also accumulate depending on their properties \cite{Ettner2024}.  
%

\subsection{Tumor and Distribution Studies} 
\label{subsec:cam-tumor}
The \ac{CAM} model has become increasingly important for tumor research over the few last years. For example, it is employed to gain more insights into tumor biology, metastasis, drug distribution, and for molecular screening \cite{Valdes2002, pion20223d, nowak2014chicken}. Besides for research on tumor growth and metastasis \cite{Miebach22}, the CAM model is also utilized for analyzing the distribution of specific molecules during the development and testing of targeted therapeutics \cite{Ettner2024, Pawlikowska2020}.

\begin{figure}[t]
    \centering
    \includegraphics[width=0.8\linewidth]{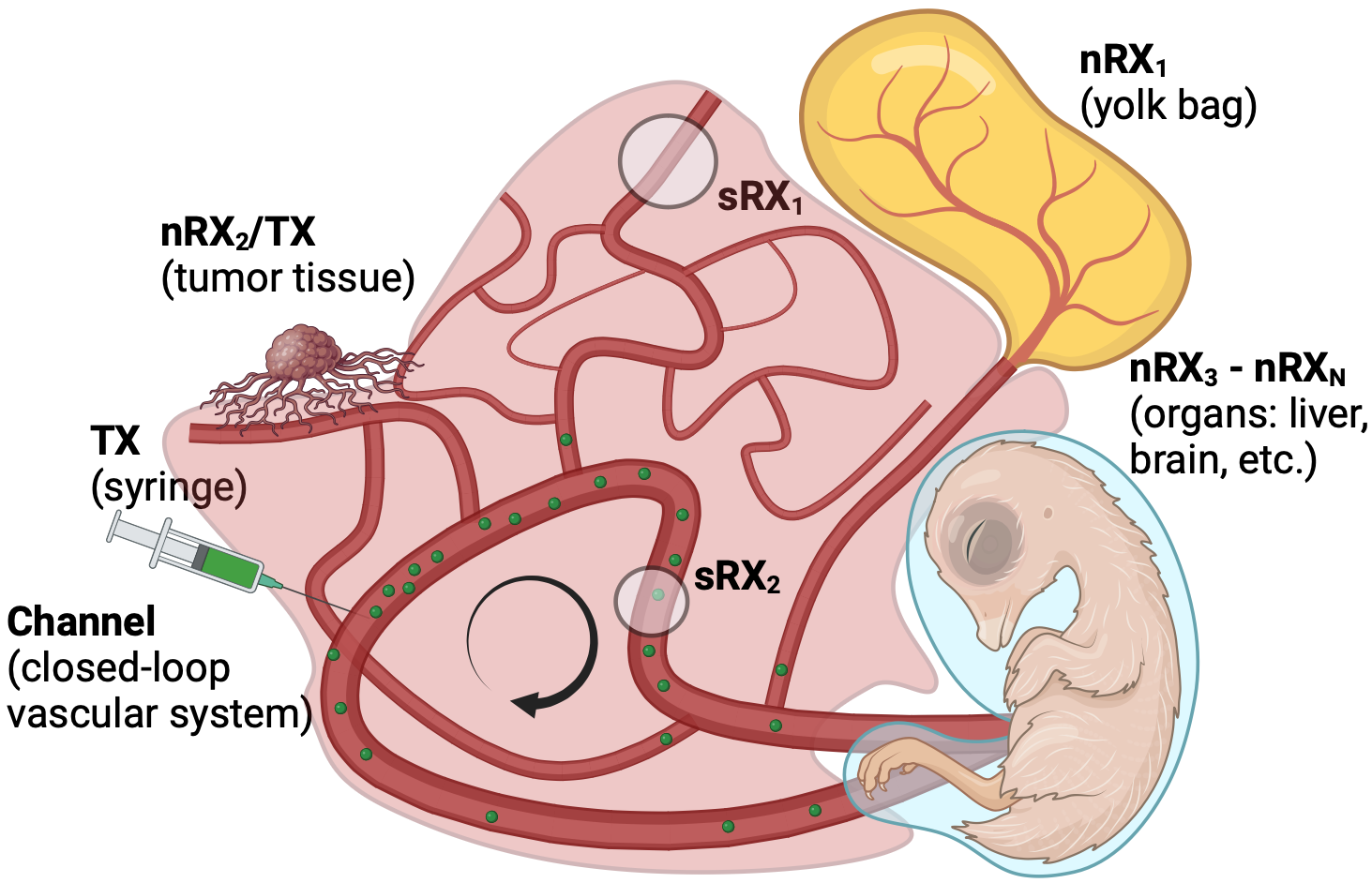}
    \vspace*{-0.4cm}
    \caption{\small Interpretation of the \ac{CAM} model as closed-loop \ac{MC} system with \acp{TX}, e.g., a syringe or a tumor, several \acp{nRX}, e.g., yolk bag, organs, tumor tissue, and several \acp{sRX} (Created with \url{BioRender.com}).}
    \label{fig:cam-mc}
    \vspace*{-0.6cm}
\end{figure}

One approach for studying the distribution and accumulation of molecules and for tumor imaging is fluorescence imaging of different types of fluorescent molecules injected into the \ac{CAM} model \cite{Zucal21, Zucal22,Ettner2024}. A fluorescent molecule, commonly used in the \ac{CAM} model, is the near-infrared fluorescent agent \acf{ICG}, which is mainly used for cancer imaging and image-guided surgeries in clinical settings \cite{Weixler23}. Stimulated with light in the near-infrared range ($750~\si{\nano\meter}$ and $950~\si{\nano\meter}$), \ac{ICG} starts to fluoresce, allowing visualization of vascular and anatomical structures (see right hand side of Fig.~\ref{fig:icg-injection}). Due to its chemical composition, \ac{ICG} interacts with membrane and macro-molecular serum proteins, including albumin, which is commonly enriched in many cancers \cite{Weixler23, Ettner2024}. 
Therefore, \ac{ICG} is often used in clinical practice for the intraoperative visualization of tumors \cite{Chauhan23}. Moreover, as \ac{ICG} preferably accumulates in tumor tissue and partly shows similar properties as drug molecules, it can be employed in the \ac{CAM} model to characterize and optimize the accumulation of molecules in tumor tissue as a basis for the design and optimization of drug delivery systems \cite{Weixler23,Chauhan23,Ettner2024}. Figure~\ref{fig:cam-drug-study} shows an example of the \ac{ICG} distribution in the vascular system of the \ac{CAM} model and its accumulation in tumor tissue \cite{Ettner2024}.

\subsection{CAM Model as an MC System}
\label{subsec:cam-mc}

Figure~\ref{fig:cam-mc} shows a representation of the \ac{CAM} model from Fig.~\ref{fig:cam-schematic} rearranged as an \ac{MC} system. The system consists of \acp{TX}, e.g., a syringe or a tumor, and possibly multiple \acp{RX} where we distinguish between \acfp{nRX}, such as the yolk bag, organs, and tumor tissue, and \acfp{sRX} such as localized measurement units. Between the \acp{TX} and \acp{RX}, the closed-loop vascular system of the \ac{CAM} represents the channel. 

\begin{figure}[t]
    \centering
    \begin{minipage}{0.49\linewidth}
        \includegraphics[width=\linewidth]{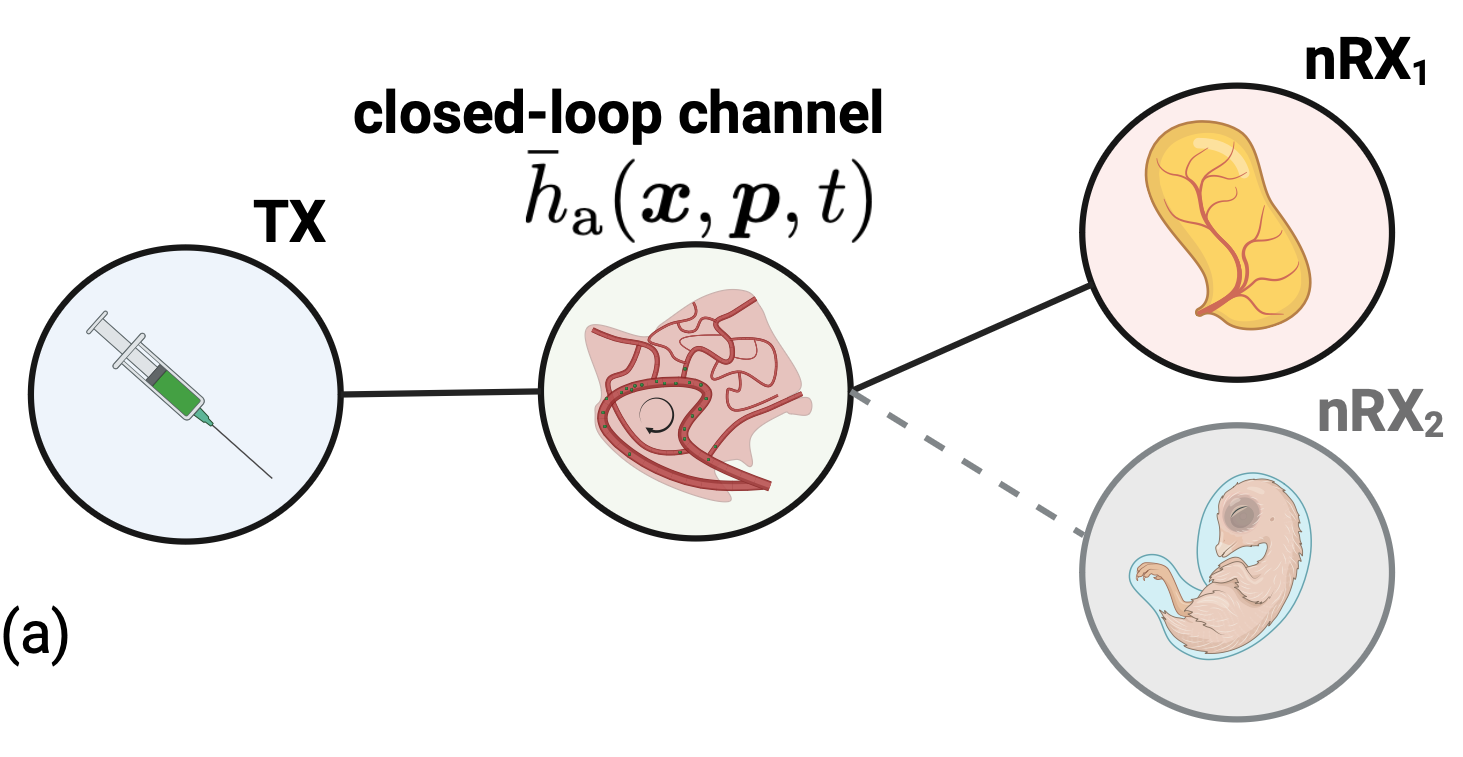}
    \end{minipage}\hfill
    \begin{minipage}{0.49\linewidth}
        \includegraphics[width=\linewidth]{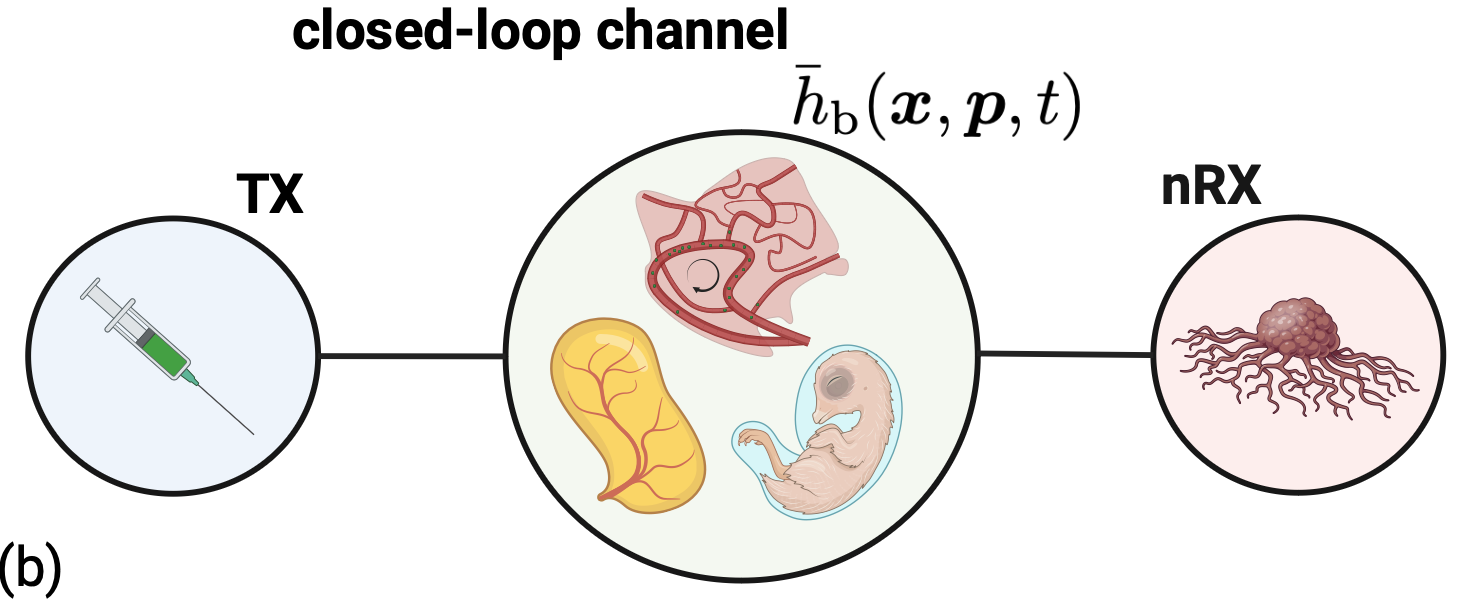}
    \end{minipage}\\
    \centering
    \includegraphics[width=0.5\linewidth]{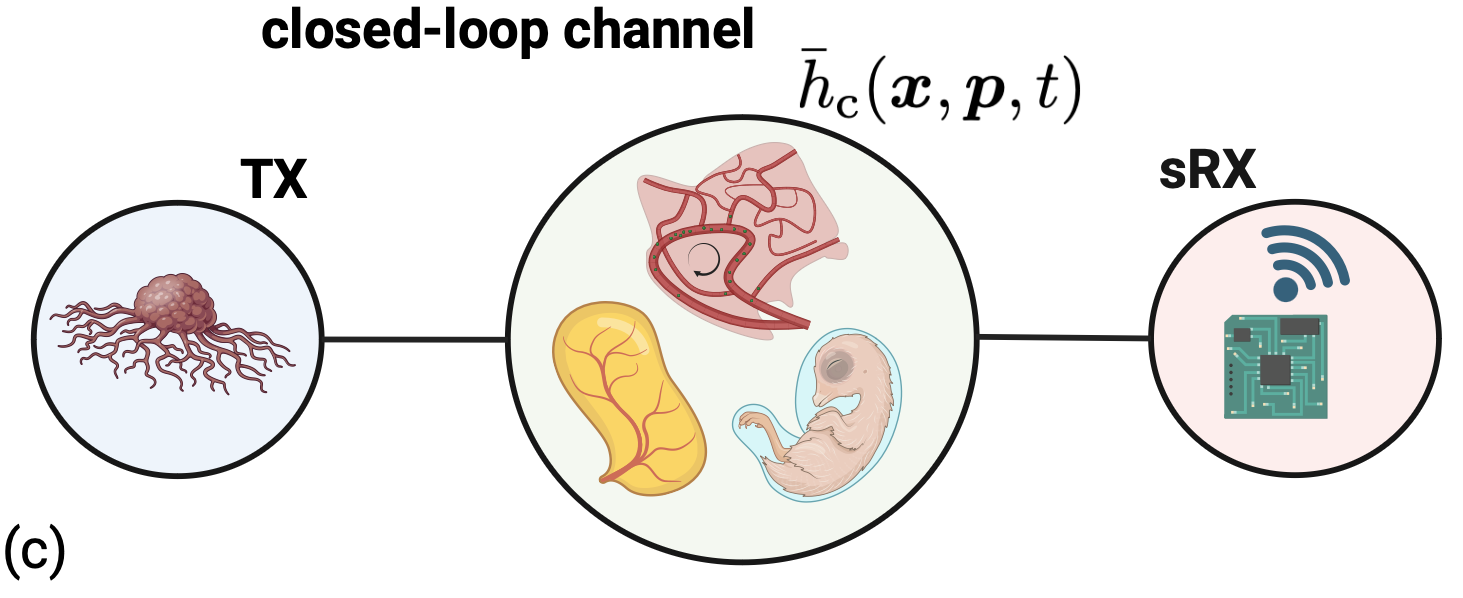}
    \vspace*{-0.4cm}
    \caption{\small Application-dependent specializations of system in Fig.~\ref{fig:cam-mc}. (a): Study molecule distribution and accumulation, yolk bag and organs are \acp{RX}. (b): Study molecule accumulation in tumors, engrafted tumor tissue is the \ac{RX}, and the yolk bag and organs are part of the channel. (c): Anomaly detection or monitoring tasks, engrafted tumor tissue is the \ac{TX} (Created with \url{BioRender.com}).}
    \label{fig:cam-mc-spec}
    \vspace*{-0.7cm}
\end{figure}

Depending on the application, the topology in Fig.~\ref{fig:cam-mc} can be specialised as it is shown for three specific cases in Fig.~\ref{fig:cam-mc-spec}. Fig.~\ref{fig:cam-mc-spec}~(a) shows the \ac{MC} system topology if the distribution and accumulation of molecules injected into the \ac{CAM} model are studied. In this scenario, depending on the \ac{DED}, the yolk bag and organs can be interpreted as multiple \acp{RX}, where molecules accumulate. The closed-loop vascular system is the channel, which can be described by a time-variant, possibly non-linear function $\bar{h}_\mathrm{a}(\bm{x},\bm{p},t)$, depending on the three-dimensional space variable $\bm{x}$, a set of environmental parameters $\bm{p}$, and time $t$. Hereby, the time-variance of the channel mainly arises from random movements of the embryo, its growth, and the non-constant blood flow. 

Figures~\ref{fig:cam-mc-spec}~(b) and (c) show ``point-to-point'' \ac{MC} systems, where a \ac{TX} releases signaling molecules for an \ac{nRX} or \ac{sRX}. Figure~\ref{fig:cam-mc-spec}~(b) shows a drug delivery system as a particular application, i.e., the \ac{TX} injects drug molecules into the \ac{CAM} model and the accumulation in the tumor tissue can be first characterized and later optimised \cite{Ettner2024}. In this scenario, the \ac{nRX} is human tumor tissue engrafted on the \ac{CAM} (see Fig.~\ref{fig:cam-drug-study}), and the organs and yolk bag are part of the closed-loop channel $\bar{h}_\mathrm{b}$ influencing the propagation of drug molecules by degradation effects. Fig.~\ref{fig:cam-mc-spec}~(c) shows a health monitoring application, where tumor tissue, the \ac{TX}, releases specific biomarkers into the closed-loop channel $\bar{h}_\mathrm{c}$, which can be detected by an \ac{sRX}. 

\section{Channel Modeling}
\label{sec:particleDis}
\begin{figure}[t]
    \centering
    \includegraphics[width=0.99\linewidth]{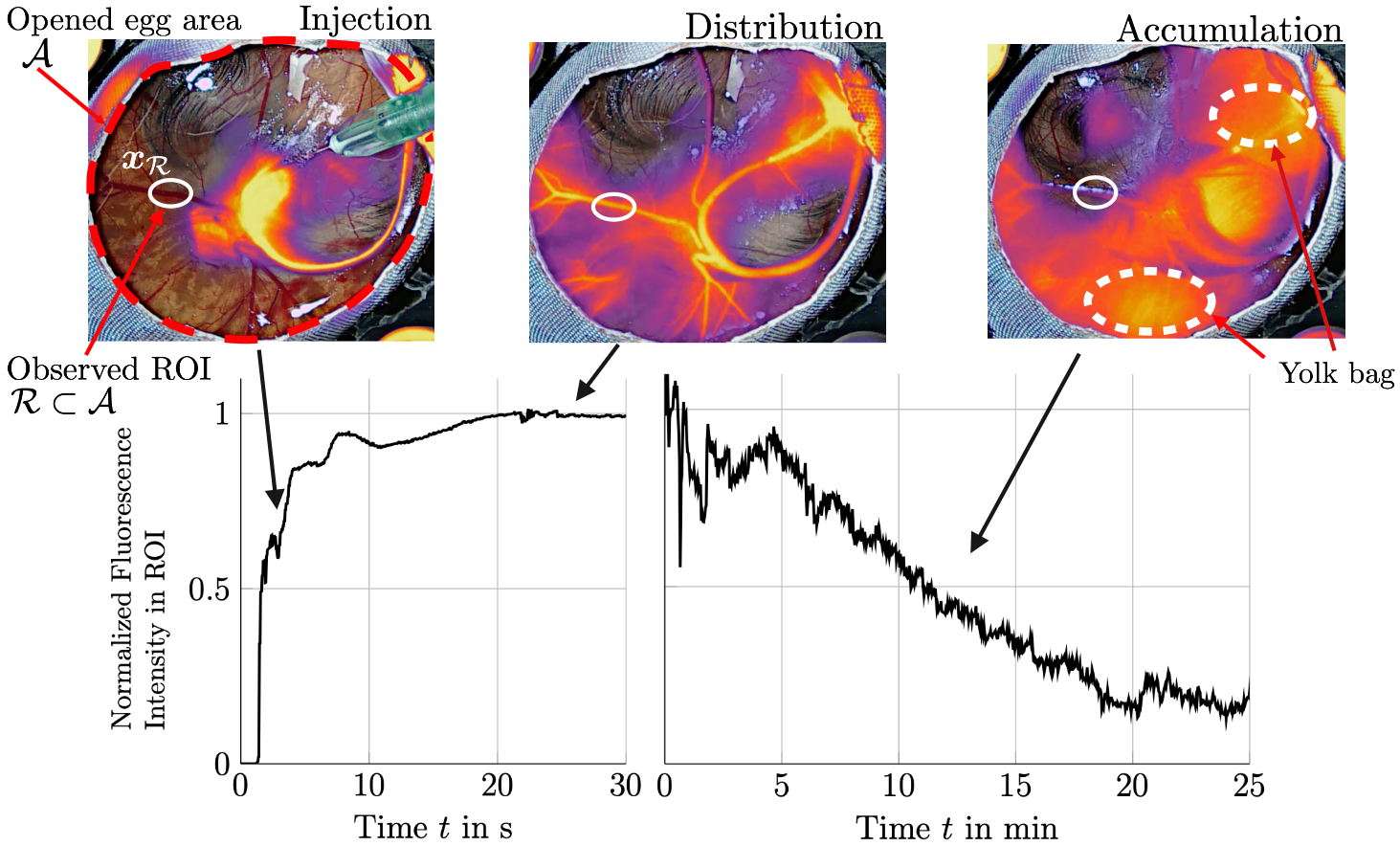}
    \vspace*{-0.2cm}
    \caption{\small Top (from left to right): Injection and distribution of \ac{ICG} molecules in the vascular system and accumulation in the yolk bag. Bottom: Measured fluorescence intensity in \ac{ROI} (Created with Biorender.com).}
    \label{fig:dist-dec-overview}
    \vspace*{-0.4cm}
\end{figure}

In this section, we consider a \ac{CAM}-based \ac{MC} system similar to Fig.~\ref{fig:cam-mc-spec}~(a), and investigate the distribution of \ac{ICG} molecules in the \ac{CAM} vascular system. In particular, the closed-loop vascular system is the channel $\bar{h}(\bm{x},\bm{p},t)$ and the injection via a syringe at $\bm{x}_{\mathrm{tx}}$ the \ac{TX}. We observe the \ac{ICG} fluorescence in a \ac{ROI} $\varcal{R}\subset\varcal{A}$ centered at $\bm{x}_\varcal{R}$, where $\varcal{A}$ is the opened egg area observed by an \ac{ICG} fluorescence camera (see Fig.~\ref{fig:dist-dec-overview}), which corresponds to a transparent \ac{sRX} located at $\bm{x}_\varcal{R}$.

The topology of the \ac{CAM} model is highly complex due to its vascularization and closed-loop character, it is constantly changing over time, and partly unknown. Therefore, the development of an exact model for the propagation of molecules is infeasible. So far, only individual vessel trees of the \ac{CAM} vascular system have been modelled explicitly and simulated numerically for hemodynamic studies \cite{Maibier2016, Kloosterman2014}. However, besides their high computational costs, such numerical models provide only limited insight into the overall dynamics and limited applicability for analytical characterization. Therefore, we propose an approximate modeling approach to characterize the molecule distribution. 
\subsection{Modeling Assumptions}
\label{subsec:assumpt}

Fig.~\ref{fig:dist-dec-overview} shows the distribution of the fluorescent molecule \ac{ICG} inside the \ac{CAM} model at different time instances (top of Fig.~\ref{fig:dist-dec-overview}) and the fluorescence measured over time in \ac{ROI} $\varcal{R}$ centered at $\bm{x}_\varcal{R}$ (bottom of Fig.~\ref{fig:dist-dec-overview}). From the figure, we make two main observations: 
\begin{itemize}
    \item[(i)] After injection, the \ac{ICG} molecule distribution reaches an equilibrium inside \ac{ROI} $\varcal{R}$ very quickly (after approx. $25~\si{\second}$). This is due to the laminar flow, dispersion, and turbulences introduced by the high vascularity and the closed-loop character of the vascular system \cite{Kloosterman2014, Maibier2016}.
    \item[(ii)] After distribution, molecules are absorbed from the vascular system into the yolk bag and organs leading to a decrease in \ac{ROI} $\varcal{R}$ (right hand side of Fig.~\ref{fig:dist-dec-overview}). Compared to the distribution, this process is much slower, i.e., in the order of minutes. 
\end{itemize}
Based on observations (i) and (ii), we neglect the absorption of molecules by the yolk bag and organs for the modeling of the injection and distribution dynamics. Therefore, we propose an approximate model for the molecule distribution inside the vascular system of the \ac{CAM} model, while the modeling of molecule absorption from the vascular system and accumulation in specific regions is left for future work. 


\subsection{Diffusion and Flow in Closed-loop Systems}

\begin{figure}[t]
    \centering
    \includegraphics[width=0.9\linewidth]{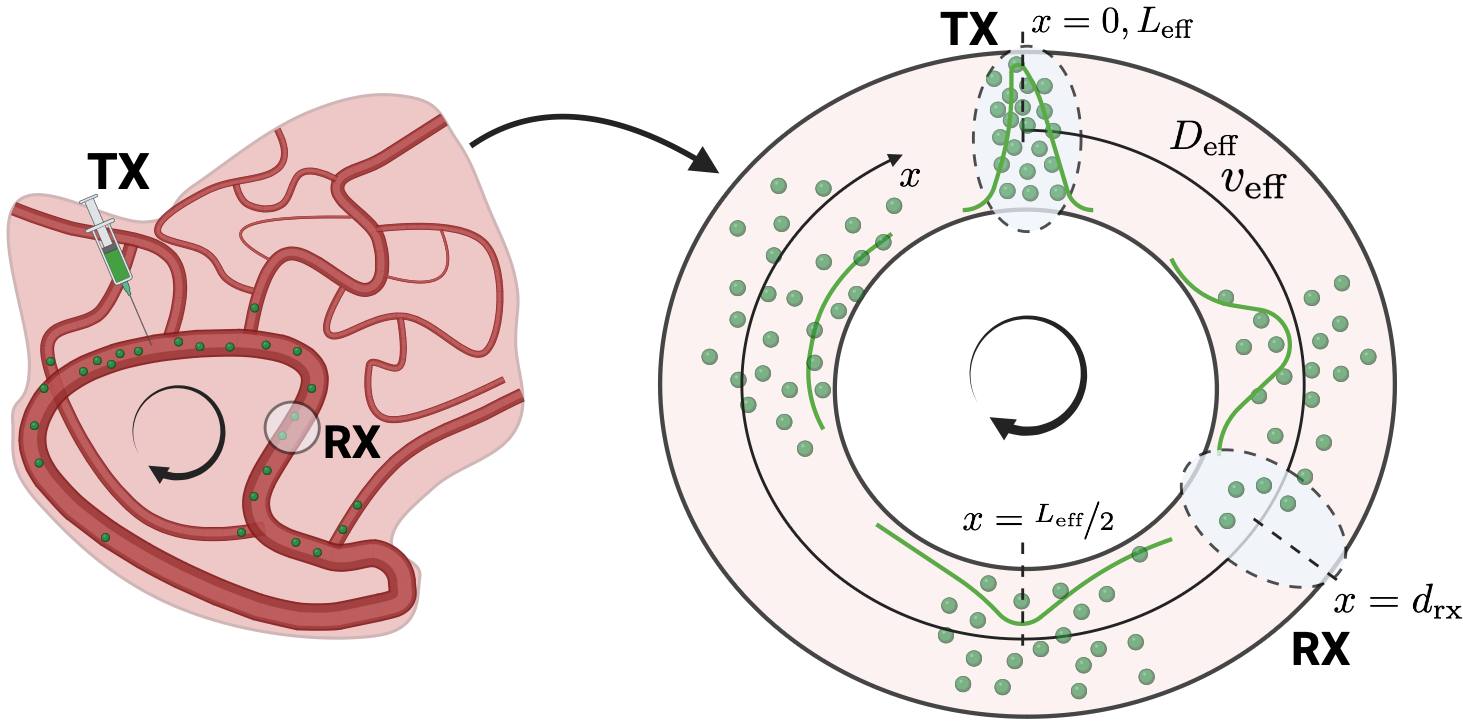}
    \caption{\small Proposed approximative model for the distribution of molecules in closed-loop systems. Left: Closed-loop \ac{CAM} vascular system. Right: Approximation as a closed-loop pipe with effective diffusion coefficient $\Deff$ and flow $\veff$ (Created with \url{BioRender.com}).}
    \label{fig:cam-mc-approx}
\end{figure}

Here, we propose an approximate model for the propagation of molecules in dispersive closed-loop systems, which we will use in Section~\ref{sec:exp} to characterize the propagation of molecules between a \ac{TX} and an \ac{RX} inside the \ac{CAM} model. 
As shown in Fig.~\ref{fig:cam-mc-approx}, we approximate the $3$D closed-loop highly vascularized system by a $3$D closed-loop pipe of length $\Leff$. In the closed-loop pipe, molecules are transported by diffusion and flow.  

Due to the strong vascularization, non-constant flow and random movements, we expect that the molecule transport inside the \ac{CAM} vascular system is highly dispersive. Therefore, we assume that molecule transport occurs in the dispersive regime \cite{Jamali2019}, and model the 3D pipe on the right hand side of Fig.~\ref{fig:cam-mc-approx} by a 1D Aris-Taylor dispersion model. In this case, we can describe the concentration of molecules $p(x,t)$ in the closed-loop pipe system in Fig.~\ref{fig:cam-mc-approx} by a $1$D drift-diffusion process \cite{Probstein2005}
\begin{align}
    \partial_t p(x,t) = \Deff \partial_{xx} p(x,t) - \veff \partial_x p(x,t), 
    \label{eq:aris}
\end{align}
where $\partial_t$ and $\partial_{xx}$ denote first order partial time- and second order space derivative, respectively, and the space coordinate $x\in[0, \Leff]$ is along the pipe (see Fig.~\ref{fig:cam-mc-approx}). Parameters $\Deff$ and $\veff$ denote the effective diffusion coefficient and effective flow velocity, respectively.


Equation~\eqref{eq:aris} can be solved straightforwardly if we assume $\Leff\to\infty$ and a uniform molecule release at $x = 0$ and $t = 0$, yielding a normal distribution in $x$ \cite{Jamali2019}
\begin{align}
    p_\mathrm{n}(x,t) = \frac{1}{\sigma(t)\sqrt{2\pi}}\exp\left(-\frac{(x - \mu(t))^2}{2\sigma^2(t)}\right),
    \label{eq:arisInfinite}
\end{align}
with variance $\sigma^2(t) = 2 \Deff t$ which is shifted along the $x$-axis with mean $\mu(t) = \veff t$. Solution \eqref{eq:arisInfinite} is commonly applied for the modeling of dispersive \ac{MC} systems in infinitely long tubes \cite{Jamali2019,Wicke2018}. However, \eqref{eq:arisInfinite} cannot be applied to closed-loop systems with finite $\Leff$, such as the \ac{CAM} model and many other envisioned application environments for \ac{MC}, e.g., the human circulatory system. 

To extend \eqref{eq:arisInfinite} to a closed-loop system with finite $\Leff$ (see right hand side of Fig.~\ref{fig:cam-mc-approx}), \eqref{eq:aris} has to be restricted to the domain $[0, \, \Leff]$ with periodic boundary conditions, i.e., $p(0,t) = p(\Leff,t)$ and $\partial_x p(0,t) = \partial_x p(\Leff,t)$. Then, \eqref{eq:aris} characterizes the $1$D dispersive propagation of molecules on a circle with circumference $\Leff$. A solution to \eqref{eq:aris} with periodic boundary conditions for a uniform release at $x =0$ and $t = 0$ follows as a wrapped normal distribution, which can be obtained from wrapping \eqref{eq:arisInfinite} to a circular domain \cite{Mardia1999}  
\begin{align}
    p_\mathrm{wn}(x,t) = \frac{1}{\sqrt{2\pi}\bar\sigma(t)}\sum_{k=-\infty}^{\infty}\exp\left(\frac{-(\bar x - \bar\mu(t) + 2\pi k)^2}{2\bar\sigma^2(t)} \right), 
    \label{eq:wrappedNormal}
\end{align}
where $\bar\sigma^2(t) = \lambda^2\sigma^2(t)$, $\bar\mu(t) = \lambda\mu(t)$ and $\bar x = \lambda x$ denote the variance, mean, and position mapped on a circle of circumference $\Leff$ with scaling parameter $\lambda = \frac{2\pi}{\Leff}$. 
The general behaviour of \eqref{eq:wrappedNormal} is depicted in Fig.~\ref{fig:cam-mc-approx}. At time $t = 0$, molecules are concentrated at $x = 0$ and start to propagate, mainly driven by flow $\veff$. Due to diffusion, variance $\bar{\sigma}^2$ is increasing over time and molecules are getting more and more dispersed. For $t \to \infty$, variance $\bar{\sigma}^2(t) \to \infty$, and molecules are uniformly distributed in the closed-loop system. In the following, we will use \eqref{eq:wrappedNormal} as an approximation for the channel characteristics $\bar{h}(\bm{x},\bm{p},t)$ of the \ac{CAM} vascular system, i.e., $\bar{h}(\bm{x},\bm{p},t) \approx p_\mathrm{wn}(x,t)$.

\begin{figure}
    \centering
    \includegraphics[width=0.9\linewidth]{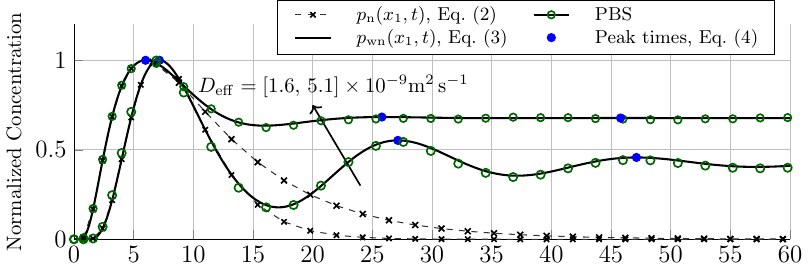}\\
    \includegraphics[width=0.9\linewidth]{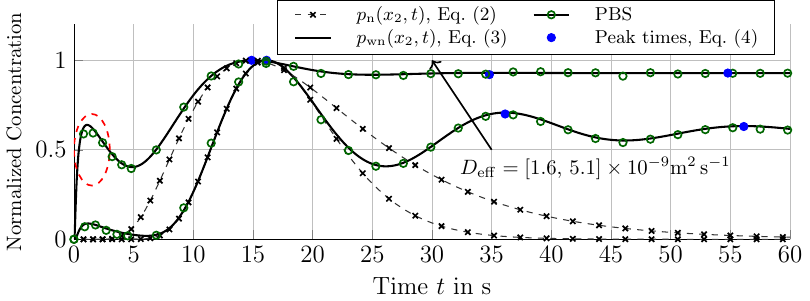}
    \vspace*{-0.35cm}
    \caption{\small Comparison between $p_\mathrm{n}$ in \eqref{eq:arisInfinite} and $p_\mathrm{wn}$ in \eqref{eq:wrappedNormal} in a closed-loop pipe at two different \ac{RX} locations $x_1$ (top) and $x_2$ (bottom) for different effective diffusion coefficients $\Deff$.}
    \label{fig:compDist}
    \vspace*{-0.3cm}
\end{figure}

Inspecting $p_\mathrm{wn}$ in \eqref{eq:wrappedNormal} and $p_\mathrm{n}$ in \eqref{eq:arisInfinite}, it becomes clear that 
\eqref{eq:arisInfinite} corresponds to the term $k = 0$ on the right hand side of \eqref{eq:wrappedNormal} for $\lambda = 1$. The other terms $k\neq 0$ in the sum in \eqref{eq:wrappedNormal} are responsible for the contributions from multiple cycles in the closed-loop. For the considered scenario in Fig.~\ref{fig:cam-mc-approx}, the closed-loop also affects the received signal, as a single release may cause multiple peaks in the average number of observed molecules at the \ac{RX}. All peak times $\{t_\mathrm{max}(k, d_\mathrm{rx}) \vert k \in \mathbb{Z}\}$ observed by the \ac{RX} located at $d_\mathrm{rx}$ (see Fig.~\ref{fig:cam-mc-approx}) due to the molecules circulating in the loop, can be obtained by modifying the peak time for an infinitely long tube \cite{Wicke2018} as
\begin{align}
    t_\mathrm{max}(k, d_\mathrm{rx}) = \frac{\Deff}{\veff^2}\left(-1 + \sqrt{1 + \frac{\veff^2}{\Deff^2} (d_\mathrm{rx} + k\,\Leff)^2} \right).
    \label{eq:peakTime}
\end{align}

\subsection{Analysis and Validation}

To analyze the properties of \eqref{eq:wrappedNormal}, we exemplary consider a micro-scale closed-loop pipe system with length $\Leff=1~\si{\milli\meter}$, radius $r_0=100~\si{\micro\meter}$,  effective flow velocity $\veff=50~\si{\micro\meter\per\second}$, and two different diffusion coefficients $D = [1.25, \, 5]\times 10^{-9}~\si{\meter\per\second}$, yielding an arrangement in the dispersive regime \cite{Schaefer2021}. The corresponding effective diffusion coefficients $\Deff$ were calculated using \cite[Eq.~(12)]{Wicke2018}. Molecules are released instantaneously at time $t = 0$ at position $x =0$. Figure~\ref{fig:compDist} shows the results of \eqref{eq:arisInfinite} and \eqref{eq:wrappedNormal} for two different \ac{RX} positions $x_1 = 0.39~\si{\milli\meter}$ and $x_2 = 0.84~\si{\milli\meter}$. The approximate peak times according to \eqref{eq:peakTime} are indicated by blue dots. Moreover, to validate \eqref{eq:wrappedNormal} and the modeling of the 3D pipe in the right hand side of Fig.~\ref{fig:compDist} by a 1D drift-diffusion process, we compare our results to the results from particle-based simulations (PBS), shown as green circles in Fig.~\ref{fig:compDist}. 

From Fig.~\ref{fig:compDist}, we observe that $p_\mathrm{wn}$ (black curves) successfully captures the effects introduced by a closed-loop system, such as the repeated observation of molecules by the \ac{RX} and the non-zero equilibrium concentration. In comparison, $p_\mathrm{n}$ (grey curve with markers) only captures the dynamics of the first peak as expected. An interesting effect, occurring in closed-loop systems is shown in the bottom plot of Fig.~\ref{fig:compDist} (highlighted by the red ellipse). For an \ac{RX} position closely behind the \ac{TX} and strong diffusion, an upstream peak occurs before the actual peak arrives at $t = 15\si{\second}$. This peak originates from molecules diffusing against the direction of flow into the \ac{RX} region. All results are in good agreement with the results from PBS, validating that \eqref{eq:wrappedNormal} is a suitable model for the propagation of molecules in dispersive closed-loop systems. 

\begin{figure}[t]
    \centering
    \includegraphics[width=0.9\linewidth]{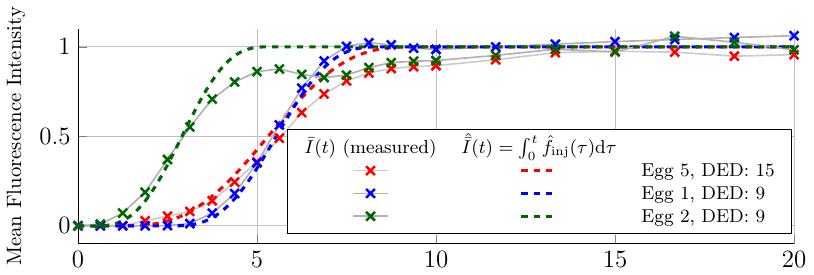}\\
    \includegraphics[width=0.9\linewidth]{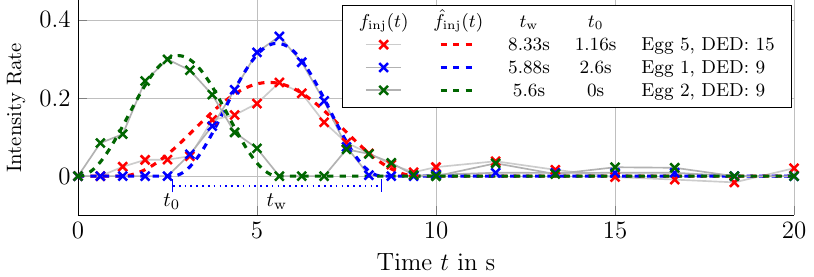}
    \vspace*{-0.4cm}
    \caption{\small Top: Measured $\bar{I}(t)$ and fitted $\hat{\bar{I}}(t)$ mean fluorescence intensity for three eggs. Bottom: Measured and fitted injection dynamics $\hat f_\mathrm{inj}(t)$ and estimated injection delays and duration.}
    \label{fig:injection_est}
    \vspace*{-0.1cm}
\end{figure}

\section{Experimental Study}
\label{sec:exp}

In this section, we present experimental results for the distribution of molecules in the \ac{CAM} model. 
For the experiments, we use the fluorescent molecule \ac{ICG} (see Section~\ref{subsec:cam-tumor}) injected via a syringe into the \ac{CAM} vascular system. We use three different eggs, one at \ac{DED} 15 (Egg 5) and two at \ac{DED} 9 (Egg 1 and Egg 2). For a more detailed description of the egg preparation, we refer the reader to Section~\ref{sec:methods} and \cite{Ettner2024}. 
We measure the fluorescence intensity $I(\bm{x},t)$ over time with an \ac{ICG} camera focusing on the opened egg area $\varcal{A}$ with area size $A$\footnote{We note that we assume that the measured \ac{ICG} fluorescence is proportional to the \ac{ICG} concentration in $\bm{x}_\varcal{R}$, i.e., $I(\bm{x}_\varcal{R},t) \propto p(\bm{x}_\varcal{R},t)$ \cite{Chon2023}.}. From the videos we extract the localized fluorescence intensity $I(\bm{x}_{\varcal{R}},t)$ in a specific \ac{ROI} $\varcal{R} \subset \varcal{A}$ centered at $\bm{x}_{\varcal{R}}$. The obtained data are first used to investigate the \ac{ICG} injection dynamics and then to fit the parameters of the proposed model \eqref{eq:wrappedNormal}, to demonstrate that it is capable of approximating $I(\bm{x}_{\varcal{R}},t)$. 



\subsection{\ac{ICG} Injection Dynamics}

To model the \ac{ICG} injection dynamics, we measure the mean fluorescence intensity $\bar{I}(t)$ in $\varcal{A}$, i.e., $\bar{I}(t) = \frac{1}{A}\int_{\varcal{A}} I(\bm{x},t)\,\mathrm{d}\bm{x}$, where $I(\bm{x},t)$ is the fluorescence intensity of a single point in $\varcal{A}$. Then, the injection dynamics can be obtained from the derivative of the mean intensity $\bar{I}(t)$, i.e., $f_\mathrm{inj}(t) = \frac{\mathrm{d}}{\mathrm{d} t} \bar{I}(t)$. 

The curves with markers in Fig.~\ref{fig:injection_est} show the measured mean fluorescence intensity $\bar{I}(t)$ (top plot) and the injection dynamics $f_\mathrm{inj}$ (bottom plot) for three different eggs. 
From the measured $f_\mathrm{inj}$, we first extract the injection duration $t_\mathrm{w}$ and delay $t_\mathrm{0}$, as indicated for Egg 1 in Fig.~\ref{fig:injection_est}. Then, we model the injection dynamics $f_\mathrm{inj}$, by a raised cosine as follows
\begin{align}
    f_\mathrm{inj}(t)\approx \hat{f}_\mathrm{inj}(t) = \nicefrac{1}{t_\mathrm{w}} \left(1 -\cos(\omega (t - t_\mathrm{0}) \right), 
    \label{eq:raised}
\end{align}
with $\omega = 2\pi / t_\mathrm{w}$. The dashed curves in the bottom plot of Fig.~\ref{fig:injection_est} show the approximation $\hat{f}_\mathrm{inj}$ of $f_\mathrm{inj}$, and the resulting mean intensity $\hat{\bar{I}}(t) = \int_0^t \hat{f}_\mathrm{inj}(\tau)\mathrm{d}\tau$ in the top plot. It can be observed that $\hat{f}_\mathrm{inj}$ from \eqref{eq:raised} is in good agreement with the measured injection dynamics $f_\mathrm{inj}$ for all eggs. Therefore, we will use $\hat{f}_\mathrm{inj}$ to account for the injection dynamics.

\begin{figure}[t]
    \centering
    \includegraphics[width=\linewidth]{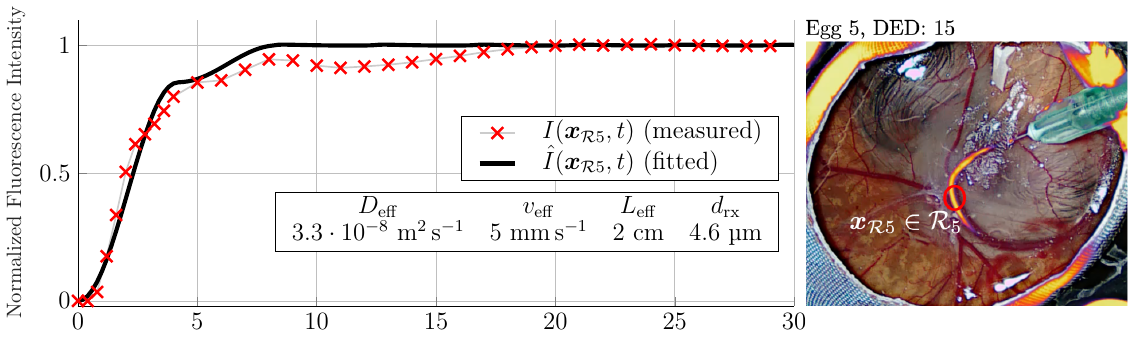}\\[-0.5em]
    \includegraphics[width=\linewidth]{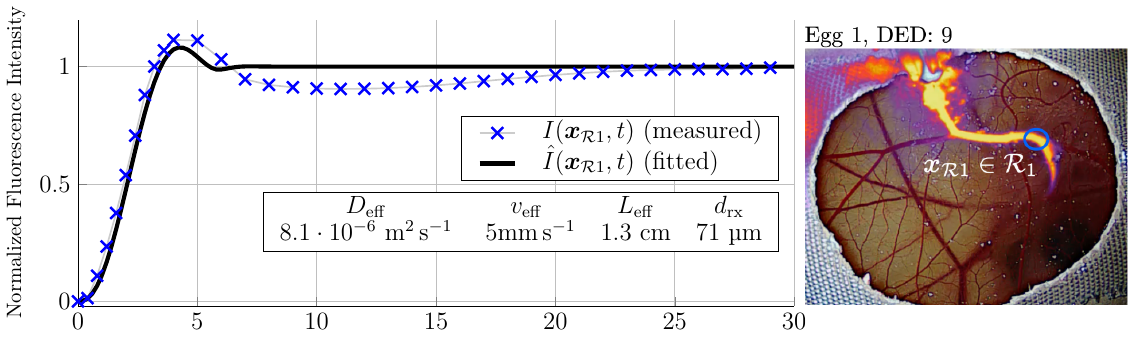}\\[-0.5em]
    \includegraphics[width=\linewidth]{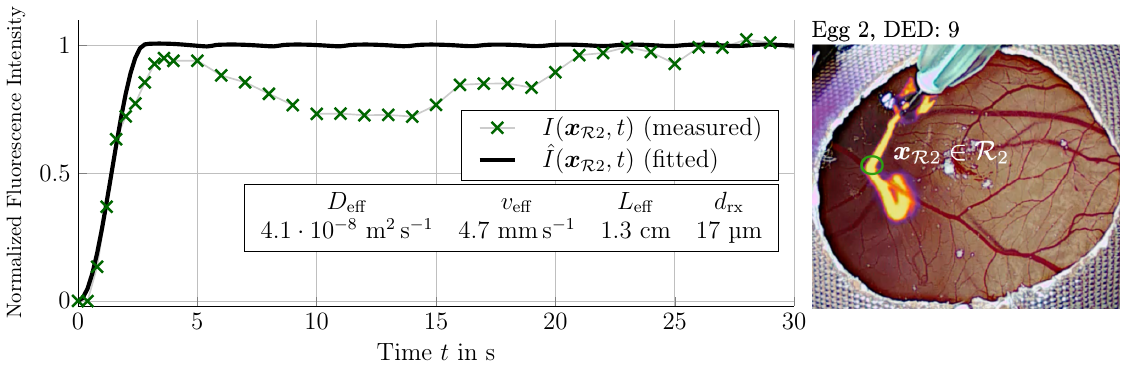}
    \vspace*{-0.7cm}
    \caption{\small Experimental and fitting results for the distribution of \ac{ICG} molecules inside the \ac{CAM} vascular system (left hand side), measured in specific \acp{ROI} (right hand side) for three different eggs.}
    \label{fig:results}
    \vspace*{-0.35cm}
\end{figure}


\vspace*{-0.2cm}
\subsection{\ac{ICG} Distribution}

In the following, we will evaluate the performance of the proposed approximate model for the molecule distribution in the \ac{CAM} model (see Fig.~\ref{fig:cam-mc-approx} and Section~\ref{sec:particleDis}). In particular, we use the estimated injection dynamics $\hat{f}_\mathrm{inj}$ together with $p_\mathrm{wn}$ in \eqref{eq:wrappedNormal}, to approximate the \ac{ICG} distribution dynamics measured at $\bm{x}_\varcal{R}$ as follows
\begin{align}
I(\bm{x}_\varcal{R},t)\approx \hat{I}(\bm{x}_\varcal{R},t) =  p_\mathrm{wn}(d_\mathrm{rx},t)\ast \hat{f}_\mathrm{inj}(t),  
\label{eq:approx}
\end{align}
where $\ast$ denotes convolution. As all parameters of the \ac{CAM} vascular system are unknown, or only given very vaguely in the literature, we estimate all parameters of $p_\mathrm{wn}$ jointly, i.e., the diffusion coefficient $\Deff$, flow velocity $\veff$, length $\Leff$ and also the distance $d_\mathrm{rx}$ from the injection site\footnote{We note that the actual distance between the injection location and $\bm{x}_\varcal{R}$ could be also measured in the recordings, but as we approximate the entire vascular system by a single tube, distance $d_\mathrm{rx}$ is not the same as in the \ac{CAM} model.}. 
For parameter estimation, we use Matlab's curve fitting toolbox with the non-linear least squares method. As the measured $I(\bm{x}_\varcal{R},t)$ reach an equilibrium very quickly, the initial slope is the most important part for the estimation process. Therefore, the actual fitting is performed on the derivatives, i.e., we determine a set of parameters $[\Deff, \veff, \Leff, d_\mathrm{rx}]$ such that $\frac{\mathrm{d}}{\mathrm{d}t}\hat{I}(\bm{x}_\varcal{R},t) \approx \frac{\mathrm{d}}{\mathrm{d}t}I(\bm{x}_\varcal{R},t)$.

Figure~\ref{fig:results} shows the measured fluorescence intensity $I(\bm{x}_\varcal{R},t)$ at $\bm{x}_\varcal{R}$ over time, for three eggs (curves with colored markers). The black curves in Fig.~\ref{fig:results} show the approximated fluorescence intensity $\hat{I}(\bm{x}_\varcal{R},t)$ obtained with \eqref{eq:approx} and the estimated parameter values for each measurement. The photos on the right hand side of Fig.~\ref{fig:results} show the actual injection locations and the locations of \acp{ROI} $\varcal{R}_5$, $\varcal{R}_1$, and $\varcal{R}_2$ in Egg 5, Egg 1, and Egg 2, respectively, shortly after the injection started. In general, all measurements are in line with the assumption that the equilibrium inside the vascular system is reached very fast (see Section~\ref{subsec:assumpt}). 

We observe from Fig.~\ref{fig:results} that the overall shape of the measurement $I(\bm{x}_\varcal{R},t)$ is well reproduced by the proposed approximation $\hat{I}(\bm{x}_\varcal{R},t)$. For example, for Egg 5 (top plot) the characteristic step in the slope around $5~\si{\second}$, for Egg 1 (center plot) the peak around $4~\si{\second}$, and for Egg 2 (bottom plot) the initial slope are reproduced very well by the proposed model. However, for all eggs, the model fails to follow the intensity drop (around $10~\si{\second}$), most significant for Egg 2 in the bottom plot (see Section~\ref{sec:discuss}). Instead, approximation $\hat I$ stays at the equilibrium. Despite this mismatch, our proposed model is capable of approximating the complex behaviour of the \ac{CAM}s vascular system for a fixed injection-measurement (\ac{TX}-\ac{RX}) arrangement. 

Next, we discuss the estimated parameter values. First, we can observe that a similar $\veff$ is obtained for all three eggs, which is also in the range of values known for the \ac{CAM} model from the literature (see Section~\ref{sec:topol}). The estimated $\Deff$ values are in the range of $10^{-8}$--$10^{-6}~\si{\square\meter\per\second}$, which exceeds the typical range of molecular diffusion coefficients by several orders of magnitude. However, as the \ac{CAM} vascular system is highly vascularized and contains several sources of turbulence (cf. Section~\ref{sec:topol}), these large values are reasonable and reflect the expected dispersion. Comparing the estimated $\Leff$ values, we observe that a similar $\Leff$ is estimated for Egg 1 (center plot) and Egg 2 (bottom plot) which are at the same \ac{DED}, while the value for Egg 5 (top plot) is larger. Although the estimated parameter values are not an estimate of the actual parameters of the \ac{CAM} model because of the high level of abstraction, the relationships between them seem meaningful. 
\vspace*{-0.2cm}
\subsection{Discussion and Future Work}
\label{sec:discuss}

We have shown that the proposed model is capable of approximating the molecule propagation inside the closed-loop \ac{CAM} vascular system. However, there are deviations from the measurement data, as the model is not able to capture all environmental effects occurring in the \ac{CAM} model. Therefore, the proposed model is only a first step towards the modeling of complex closed-loop vascular systems and has to be extended further. In the following, we shortly discuss reasons for the deviations, which also motivate the most important topics for future work: 
\vspace*{-0.1cm}
\begin{itemize}
    \item The measurements of the camera are 2D representations of a 3D system. Therefore, the measurement at $\bm{x}_\varcal{R}$ can be interfered with the fluorescence of the underlying vessels. Moreover, the measurements can be strongly influenced by random movements of the embryo. 
    \item The model does not account for chemical reactions that may occur between the \ac{ICG} molecules and other molecules in the environment. Moreover, the accumulation in the yolk bag, organs, and vessel walls is not considered.
    \item The flow velocity $\veff$ in \eqref{eq:wrappedNormal} is assumed to be constant. However, the blood flow velocity inside the \ac{CAM} model is not constant but time-variant and has to be taken into account. 
\end{itemize}
Besides these points, which we plan to address in our future work, the proposed model was only evaluated for one fixed injection and measurement arrangement. In future work, we will investigate, how the estimated parameters fit for different arrangements in the same egg and between several eggs of the same \ac{DED}. 

\section{Conclusion}
\label{sec:conc}

In this paper, we introduced the \ac{CAM} model as a versatile \textit{in vivo} \ac{MC} testbed, as an important step to enable the transition from proof-of-concept  systems towards practical applications. The \ac{CAM} model consist of a closed-loop cardiovascular system including blood circulation and organs, and therefore provides a realistic environment for the validation and optimization of \ac{MC} technologies.  
We introduced an analytical model for diffusion and flow of molecules in dispersive closed-loop pipe systems and validated it by particle-based simulations. Then, the model was applied to approximate the distribution of molecules inside the \ac{CAM} model for a given injection-measurement arrangement for three different eggs. Our results confirm that the proposed model is able to approximate the molecule distribution inside the \ac{CAM} model and the estimated set of parameters values is in a physically plausible range. The existing deviations from the measured behavior are due to the fact that not all relevant effects were taken into account in the analytical model. In future work, we will extend the model to account for additional environmental effects occurring in the \ac{CAM} model such as the time-variant blood flow velocity and the accumulation of molecules in the yolk bag and organs. 

\section{Methods}
\label{sec:methods}
For all experiments, chicken eggs were incubated at constant temperature of $37.8~\si{\celsius}$, a pCO$_2$ of $5~\si{\percent}$, and humidity calibrated to $63~\si{\percent}$. They were rotated during the first four days of incubation to prevent embryos from sticking to the shell membranes. The eggshells were opened on \ac{DED} 4, because of the fusion of the \ac{CAM} and the risk of rupturing or damaging the vasculature \cite{nowak2014chicken}. Therefore, the air entrapment in the egg was localized with a light source and a hole of approximately $2~\si{\milli\meter} \times 2~\si{\milli\meter}$ was cut with sterile tweeters into the overlaying eggshell to introduce atmospheric negative pressure. A second hole of approximately $5 \si{\milli\meter} \times 5 \si{\milli\meter}$ was introduced at the longitudinal side of the egg. The holes were covered with Leukosilk® (BSN medical, Hamburg, Germany) and the eggs were returned to the incubator in a static position.
The video recordings of the \ac{ICG} molecule distribution were performed with an \ac{ICG} camera (EleVision\textsuperscript{TM} IR Platform, Medtronic).
The video data analysis (Fluorescence tracking in \acp{ROI}) was done with the Matlab Fluorescence Tracker App\footnote{\url{https://matlab.mathworks.com/open/github/v1?repo=mathworks/Fluorescence-Tracker-App}}.


\printbibliography

\end{document}